\documentclass[epj,fleqn]{webofc}
\usepackage[utf8]{inputenc}
\usepackage[varg]{txfonts}   
\usepackage{booktabs}
\usepackage{xcolor}
\definecolor{darkred}{rgb}{0.4,0.0,0.0}
\definecolor{darkgreen}{rgb}{0.0,0.4,0.0}
\definecolor{darkblue}{rgb}{0.0,0.0,0.4}
\usepackage[bookmarks,linktocpage,colorlinks,
    linkcolor = darkred,
    urlcolor  = darkblue,
    citecolor = darkgreen]{hyperref}

\usepackage{multirow}
\usepackage{bigdelim}
\usepackage{tikz} 
\usetikzlibrary{arrows}

\renewcommand{\Im}[0]{\operatorname{Im}}
\renewcommand{\d}{\mathrm{d}}
\newcommand{\tr}{\mathrm{tr}\,}

\newcommand{\Qone}{\texttt{QCD+QED1}}
\newcommand{\Sone}{\texttt{QCD+QED2}}

\wocname{EPJ Web of Conferences}
\woctitle{Lattice2017}

\begin{document}
%
\selectlanguage{english}
\title{%
\texttt{openQ*D} simulation code for QCD+QED
}
\author{%
\firstname{Isabel}   \lastname{Campos}\inst{1,2} \and
\firstname{Patrick}  \lastname{Fritzsch}\inst{3} \and
\firstname{Martin}   \lastname{Hansen}\inst{4} \and
\firstname{Marina}   \lastname{Krsti\'c Marinkovi\'c}\inst{3,5}\fnsep\thanks{Speaker, \email{marina.marinkovic@cern.ch}} \and \\
\firstname{Agostino} \lastname{Patella}\inst{3,6}\fnsep\thanks{Speaker, \email{agostino.patella@cern.ch}} \and
\firstname{Alberto}  \lastname{Ramos}\inst{3} \and
\firstname{Nazario}  \lastname{Tantalo}\inst{7}
}
\institute{%
Instituto de Física de Cantabria - IFCA-CSIC,
Avda. de Los Castros s/n, 39005 Santander, Spain
\smallskip \and
Instituto de Física Teórica UAM/CSIC, Universidad Autónoma de Madrid,\\
C/ Nicolás Cabrera 13-15, Cantoblanco, Madrid 28049
\smallskip \and
Theoretical Physics Department, CERN,
CH-1211 Geneva 23, Switzerland\footnote{CERN-TH-2017-217}
\smallskip \and
CP3-Origins, University of Southern Denmark,
Campusvej 55, DK-5230 Odense M, Denmark
\smallskip \and
School of Mathematics, Trinity College Dublin,
Dublin 2, Ireland
\smallskip \and
Centre for Mathematical Sciences, Plymouth University,
Plymouth, PL4 8AA, UK
\smallskip \and
Universit\`a di Roma Tor Vergata, INFN, Sezione di Tor Vergata, c/o Dipartimento di Fisica,\\
Via della Ricerca Scientifica 1, I-00133 Rome, Italy
}
\abstract{%
The \texttt{openQ*D} code for the simulation of QCD+QED with C$^\star$ boundary conditions is
presented. This code is based on \texttt{openQCD-1.6}, from which it inherits
the core features that ensure its efficiency: the locally-deflated
SAP-preconditioned GCR solver, the twisted-mass frequency splitting of the
fermion action, the multilevel integrator, the 4th order OMF integrator, the
SSE/AVX intrinsics, etc. The photon field is treated as fully dynamical and C$^\star$
boundary conditions can be chosen in the spatial directions.
We discuss the main features of \texttt{openQ*D},
and we show basic test results and performance analysis. An alpha version of this code is publicly available and can be downloaded from \url{http://rcstar.web.cern.ch/}.
}
\maketitle



\tableofcontents

\section{Introduction}
\label{intro}

The calculation of isospin--breaking corrections to hadronic observables from lattice simulations requires a theoretically sound definition of charged--hadron states in finite volume. As discussed in ref.~\cite{Lucini:2015hfa}, a possible way to preserve locality and gauge--invariance is to employ C$^\star$ boundary conditions in the space directions.

We present here the \texttt{openQ*D} package, which can be used to simulate QCD+QED or QCD in isolation with C$^\star$ boundary conditions. In the context of isospin--breaking corrections, the simulation of QCD in isolation is useful within the framework of the RM123 method~\cite{deDivitiis:2013xla}, in which observables are calculated order--by--order in the electromagnetic coupling.

An alpha version (\texttt{openQ*D-0.9a2}) is publicly available and can be downloaded from~\url{http://rcstar.web.cern.ch/}. This version of the code has already passed a large number of tests, performed by means of check programs that are provided along with the code. On top of this, several core featurs of \texttt{openQ*D-0.9a2} have been compared with an independently developed code based on \texttt{HiRep} and presented by one of the authors in another talk of this conference~\cite{hansen:lat2017}. We are currently working towards a fully-tested and stable release, which we plan to make available before the end of the year.

The general features of the \texttt{openQ*D} code are discussed in section~\ref{sec:openQxD}. In sections~\ref{sec:qcd} and~\ref{sec:qed} we discuss some exploratory runs, and selected issues related to simulations with C$^\star$ boundary conditions, with or without dynamical U(1) field.

\section{\texttt{openQ*D} code: general features}
\label{sec:openQxD}

\texttt{openQ*D} is an extension of the \texttt{openQCD} code~\cite{openQCD,Luscher:2012av}, from which it inherits the core features, and most notably the Dirac operator and the solvers. The \texttt{openQ*D-0.9a2} code supports:
\begin{itemize}
   
   \item Simulation of the SU(3) gauge theory (\textit{inherited}) and the SU(3)$\times$U(1) gauge theory (\textit{extension}).
   
   \item A one--parameter family of SU(3) gauge actions, built with plaquettes and planar double--plaquettes. This family includes the Wilson, L\"uscher--Weisz~\cite{Luscher:1984xn} and Iwasaki~\cite{Iwasaki:2011np} actions (\textit{inherited}).
   
   \item A one--parameter family of U(1) gauge actions, built with plaquettes and planar double--plaquettes (\textit{extension}).
   
   \item $O(a)$-improved Wilson quarks~\cite{Sheikholeslami:1985ij} in the fundamental representation of the SU(3) gauge group (\textit{inherited}) and generic electric charge (\textit{extension}).
   
   \item Open, SF, open--SF, periodic boundary condition in time (\textit{inherited}).
   
   \item Periodic (possibly $\theta$-periodic for fermions, \textit{inherited}) or C$^\star$ boundary conditions~\cite{Wiese:1991ku} in space (\textit{extension}).
   
   \item Nested hierarchical integrators~\cite{Sexton:1992nu} for the molecular--dynamics equations, based on any combination of the leapfrog, 2nd order Omelyan--Mryglod--Folk (OMF) and 4th order OMF elementary integrators~\cite{OMELYAN2003272} (\textit{inherited}).

   \item HMC algorithm with twisted--mass Hasenbusch frequency splitting~\cite{Hasenbusch:2001ne,Hasenbusch:2002ai}. Optionally with even--odd preconditioning~\cite{DeGrand:1988vx} (\textit{inherited}).
   
   \item RHMC algorithm~\cite{Kennedy:1998cu} with frequency splitting and even--odd preconditioning (\textit{inherited}).
   
   \item Rational approximation to a generic fractional power of the fermion determinant with or without twisted--mass reweighting (\textit{extension}).
   
   \item Deflation acceleration and chronological solver along the molecular--dynamics trajectories~\cite{Luscher:2007se,Luscher:2007es} (\textit{inherited}).
   
   \item A choice of solvers (including multi--shift conjugate gradient and highly optimized deflated solvers) for the Dirac equation, separately configurable for each force component and pseudofermion action (\textit{inherited}).
   
   \item SSE/AVX acceleration (\textit{inherited}).
\end{itemize}

In future versions of the code, a non--compact U(1) action and the Fourier acceleration for the U(1) field will be included. In this section the distinctive features of the \texttt{openQ*D} code will be described in some detail.

\subsection{C$^\star$ boundary conditions}
\label{subsec:cstar}

\texttt{openQ*D} can simulate C$^\star$ boundary conditions in one or more space
directions, by means of an orbifold construction. In short, we say that a space direction $k$ is C--direction (resp. P--direction), if fields satisfy C$^\star$ (resp. periodic) boundary conditions along the direction $k$.

Consider a $L_0 \times L_1 \times L_2 \times L_3$ lattice, which is referred to as \textit{physical lattice}, with C$^\star$ boundary conditions along direction 1 and possibly along directions 2 and 3. The code simulates an $L_0 \times (2L_1) \times L_2 \times L_3$ lattice, which is referred to as \textit{extended lattice}. The sets of points of the extended and physical lattice are denoted by $\Lambda_\text{ext}$ and $\Lambda_\text{phs}$ respectively, i.e.
\begin{gather}
   \Lambda_\text{ext} = \{ x \in \mathbb{Z}^4 \ | \ 0 \le x_\mu < L_\mu \text{ for } \mu \neq 1, \ 0 \le x_1 < 2L_1 \} \ , \\
   \Lambda_\text{phs} = \{ x \in \mathbb{Z}^4 \ | \ 0 \le x_\mu < L_\mu \} \ .
\end{gather}
The physical lattice is identified as a subset of the extended lattice. The set of points $\Lambda_\text{ext} \setminus \Lambda_\text{phs}$ is referred to as the \textit{mirror lattice}.

The fundamental fields in the \texttt{openQ*D-0.9a2} code are the SU(3) link variable $U(x,\mu)$ and the real photon field $A(x,\mu)$. Only the compact formulation of QED is considered here, therefore all observables are written in terms of the U(1) link variable
\begin{gather}
   z(x,\mu) = \exp \{ i A(x,\mu) \} \ ,
\end{gather}
and the real photon field can be restricted to $-\pi \le A(x,\mu) < \pi$. C$^\star$ boundary conditions along direction 1 on the physical lattice are given by
\begin{subequations}
   \label{eq:orbi}
   \begin{alignat}{2}
      &
      U(x + L_1 \hat{e}_1,\mu) = U(x,\mu)^* \ ,
      &\qquad&
      A(x + L_1 \hat{e}_1,\mu) = -A(x,\mu) \ ,
      \label{eq:orbi:gauge} \\
      &
      \psi(x + L_1 \hat{e}_1) = C^{-1} \bar{\psi}^T(x) \ ,
      &\qquad&
      \bar{\psi}(x + L_1 \hat{e}_1) = - \psi^T(x) C \ .
      \label{eq:orbi:fermions}
   \end{alignat}
\end{subequations}
The charge--conjugation matrix $C$ satisfies the following conditions
\begin{gather}
   C^T = -C \ , \qquad
   C^\dag = C^{-1} \ , \qquad
   C \gamma_\mu C^{-1} = - \gamma_\mu^T \ .
\end{gather}
On the extended lattice, points $x$ and $x + L_1 \hat{e}_1$ do not coincide, so eqs.~\eqref{eq:orbi} have to be interpreted as constraints which defines the \textit{admissible} gauge and fermion fields. Eqs.~\eqref{eq:orbi} are referred to as the \textit{orbifold constraints}.

Admissible gauge fields in the mirror lattice are completely determined by the value of the gauge field in the physical lattice via eqs.~\eqref{eq:orbi:gauge}. The integration measure over the manifold of admissible gauge fields is given by
\begin{gather}
   \label{eq:meas:gauge}
   [ \d U ]_{\Lambda_\text{phs}} = \prod_{\mu=0}^3 \prod_{x \in \Lambda_\text{phs}} \d U(x,\mu) \ ,
   \qquad
   [ \d A ]_{\Lambda_\text{phs}} = \prod_{\mu=0}^3 \prod_{x \in \Lambda_\text{phs}} \d A(x,\mu) \ ,
\end{gather}
where the products are restricted over the physical lattice. The orbifold constraint has a slightly different meaning for fermion fields. On the physical lattice $\psi$ and $\bar{\psi}$ are independent Grassmanian variables. On the extended lattice one can choose the value of $\psi$ in each point as a complete set of independent variables. The field $\bar{\psi}$ on the extended lattice is completely determined as a function of the field $\psi$ via eqs.~\eqref{eq:orbi:fermions}. By introducing the translation operator $\mathcal{T}$ as
\begin{gather}
   (\mathcal{T} \phi)(x) =
   \begin{cases}
      \phi(x + L_1 \hat{e}_1)  \qquad & \text{if } x \in \Lambda_\text{phs} \\
      \phi(x - L_1 \hat{e}_1)  \qquad & \text{if } x \in \Lambda_\text{ext} \setminus \Lambda_\text{phs}
   \end{cases}
   \label{eq:translation}
   \ ,
\end{gather}
the relation between $\psi$ and $\bar{\psi}$ can be conveniently rewritten as
\begin{gather}
   \bar{\psi} = - \psi^T C \mathcal{T} \ .
   \label{eq:majorana}
\end{gather}
The integration measure for the fermion field is given by
\begin{gather}
   [ \d \psi ]_{\Lambda_\text{phs}} [ \d \bar{\psi} ]_{\Lambda_\text{phs}}
   =
   \prod_{x \in \Lambda_\text{phs}} \d \psi(x) \d \bar{\psi}(x)
   =
   \prod_{x \in \Lambda_\text{ext}} \d \psi(x)
   =
   [ \d \psi ]_{\Lambda_\text{ext}}
   \ .
   \label{eq:meas:fermion}
\end{gather}

\begin{figure}

\centering

\begin{tikzpicture}[scale=.60]

   \draw[->] (-1.5,-1) -- ++(3,0) node[anchor=north] {1};
   \draw[->] (-1,-1.5) -- ++(0,3) node[anchor=east,text width=12mm,align=right] {C$^\star$ dir.\\$k \neq 1$};
   
   \foreach \x in {0,1,...,5} {
      \foreach \y in {0,1,...,5} {
         \draw[fill=black] (\x,\y) circle (.1);
         \draw[thick] (\x,\y) -- ++(1,0);
         \draw[thick] (\x,\y) -- ++(0,1);
      }
   }
   \foreach \x in {6,7,...,11} {
      \foreach \y in {0,1,...,5} {
         \draw[black!40,fill=black!40] (\x,\y) circle (.1);
         \draw[thick,black!40] (\x,\y) -- ++(1,0);
         \draw[thick,black!40] (\x,\y) -- ++(0,1);
      }
   }
   
   \foreach \x in {0,1,...,5} {
      \draw[red,thick] (\x,6) circle (.1);
   }
   \draw[red,thick] (-.3,5.7) rectangle (5.3,6.3);
   \draw[red,thick] (5.7,-.3) rectangle (11.3,.3);
   
   \foreach \x in {6,7,...,11} {
      \draw[green!70!black,thick] (\x,6) circle (.1);
   }
   \draw[green!70!black,thick] (5.7,5.7) rectangle (11.3,6.3);
   \draw[green!70!black,thick] (-.3,-.3) rectangle (5.3,.3);
   
   \foreach \y in {0,1,...,5} {
      \draw[blue,thick] (12,\y) circle (.1);
   }
   \draw[blue,thick] (11.6,-.4) rectangle (12.4,5.4);
   \draw[blue,thick] (-.4,-.4) rectangle (.4,5.4);
   
   \begin{scope}[yshift=-9cm]
      \draw[->] (-1.5,-1) -- ++(3,0) node[anchor=north] {1};
      \draw[->] (-1,-1.5) -- ++(0,3) node[anchor=east,text width=12mm,align=right] {P dir.\\$k \neq 1$};
      
      \foreach \x in {0,1,...,5} {
         \foreach \y in {0,1,...,5} {
            \draw[fill=black] (\x,\y) circle (.1);
            \draw[thick] (\x,\y) -- ++(1,0);
            \draw[thick] (\x,\y) -- ++(0,1);
         }
      }
      \foreach \x in {6,7,...,11} {
         \foreach \y in {0,1,...,5} {
            \draw[black!40,fill=black!40] (\x,\y) circle (.1);
            \draw[thick,black!40] (\x,\y) -- ++(1,0);
            \draw[thick,black!40] (\x,\y) -- ++(0,1);
         }
      }
      
      \foreach \x in {0,1,...,5} {
         \draw[red,thick] (\x,6) circle (.1);
      }
      \draw[red,thick] (-.3,5.7) rectangle (5.3,6.3);
      \draw[red,thick] (-.3,-.3) rectangle (5.3,.3);
      
      \foreach \x in {6,7,...,11} {
         \draw[green!70!black,thick] (\x,6) circle (.1);
      }
      \draw[green!70!black,thick] (5.7,5.7) rectangle (11.3,6.3);
      \draw[green!70!black,thick] (5.7,-.3) rectangle (11.3,.3);
      
      \foreach \y in {0,1,...,5} {
         \draw[blue,thick] (12,\y) circle (.1);
      }
      \draw[blue,thick] (11.6,-.4) rectangle (12.4,5.4);
      \draw[blue,thick] (-.4,-.4) rectangle (.4,5.4);
   \end{scope}
   
\end{tikzpicture}

\vspace*{-4mm}

\caption{\small\textbf{Global geometry of extended lattice.} The \textit{top diagram} represents a section of the extended lattice along a $(1,k)$ plane where $k=2,3$ is a C$^\star$ direction. All fields are periodic along the extended direction 1. C$^\star$ boundary conditions in the direction $k=2,3$ are replaced by shifted boundary conditions in the extended lattice. Shifted boundary conditions are imposed by properly defining the nearest neighbours of boundary sites. Empty circles in the red (resp. green, blue) rectangle have to be identified with the corresponding solid circles in the red (resp. green, blue) rectangle. The \textit{bottom diagram} represents a section of the extended lattice along a $(1,k)$ plane where $k=2,3$ is a periodic direction. In \textit{both diagrams}, the black circles represent the sites of the physical lattice, and the grey circles represent the sites of the mirror lattice.
\label{fig:geometry}
}

\vspace*{-4mm}
\end{figure}
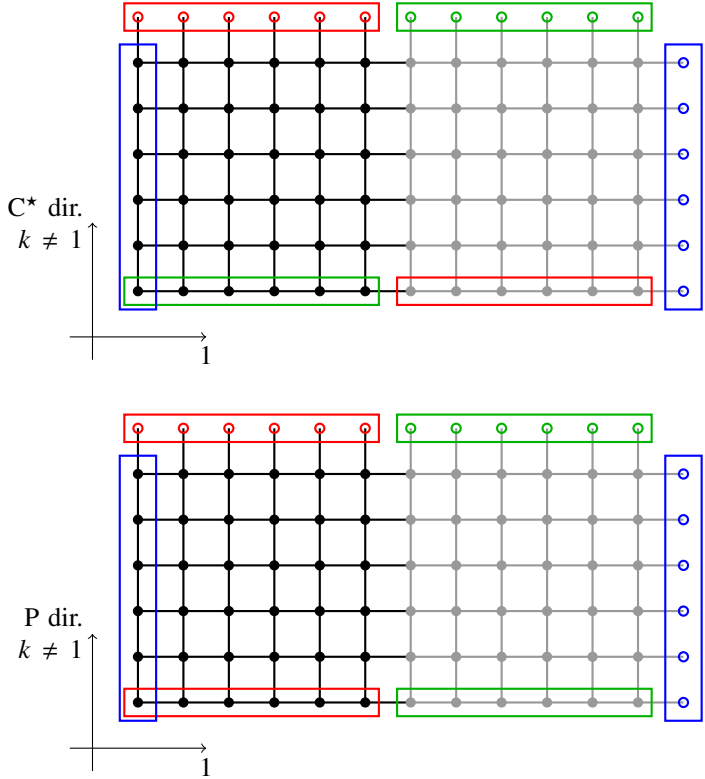

Since the square of the charge--conjugation operation is the identity, all fields must obey periodic boundary conditions along the extended direction 1, i.e.
\begin{subequations}
   \label{eq:periodic}
   \begin{alignat}{2}
      &
      U(x + 2L_1 \hat{e}_1,\mu) = U(x,\mu) \ ,
      &\qquad&
      A(x + 2L_1 \hat{e}_1,\mu) = A(x,\mu) \ ,
      \label{eq:periodic:gauge} \\
      &
      \psi(x + 2L_1 \hat{e}_1) = \psi(x) \ ,
      &\qquad&
      \bar{\psi}(x + 2L_1 \hat{e}_1) = \bar{\psi}(x) \ .
      \label{eq:periodic:fermions}
   \end{alignat}
\end{subequations}
C$^\star$ boundary conditions in directions $k=2,3$ are implemented by modifying the global geometry of the torus. If $k=2,3$ is a C$^\star$ direction, then shifted boundary conditions (see fig.~\ref{fig:geometry}) are imposed
\begin{subequations}
   \label{eq:shifted}
   \begin{alignat}{2}
      &
      U(x + L_k \hat{e}_k,\mu) = U(x + L_1 \hat{e}_1,\mu) \ ,
      &\qquad&
      A(x + L_k \hat{e}_k,\mu) = A(x + L_1 \hat{e}_1,\mu) \ ,
      \label{eq:shifted:gauge} \\
      &
      \psi(x + L_k \hat{e}_k) = \psi(x + L_1 \hat{e}_1) \ ,
      &\qquad&
      \bar{\psi}(x + L_k \hat{e}_k) = \bar{\psi}(x + L_1 \hat{e}_1) \ .
      \label{eq:shifted:fermions}
   \end{alignat}
\end{subequations}
When combined with the orbifold constraint~\eqref{eq:orbi}, shifted boundary conditions are equivalent to C$^\star$ boundary conditions in direction $k=2,3$. For instance for the SU(3) gauge field,
\begin{gather}
   U(x + L_k \hat{e}_k,\mu) = U(x + L_1 \hat{e}_1,\mu) = U(x,\mu)^* \ .
\end{gather}

\subsection{Gauge actions}
\label{subsec:gauge}

For simplicity, periodic boundary conditions in the time direction will be assumed. The SU(3) and compact U(1) gauge actions are respectively:
\begin{gather}
   S_{\text{g,SU(3)}} = \frac{\omega_{\text{C}^\star}}{g_0^2} \sum_{k=0}^1 c^\text{SU(3)}_k \sum_{\mathcal{C} \in \mathcal{S}_k} \tr [ 1 - U(\mathcal{C}) ] \ ,
   \label{eq:action:SU3} \\
   S_{\text{g,U(1)}} = \frac{\omega_{\text{C}^\star}}{2 q_\text{el}^2 e_0^2} \sum_{k=0}^1 c^\text{U(1)}_k \sum_{\mathcal{C} \in \mathcal{S}_k} [ 1 - z(\mathcal{C}) ] \ ,
   \label{eq:action:U1}
\end{gather}

Given a path $\mathcal{C}$ on the lattice, $U(\mathcal{C})$ and $z(\mathcal{C})$ denote the SU(3) and U(1) parallel transports along $\mathcal{C}$. $\mathcal{S}_0$ and $\mathcal{S}_1$ are the sets of all oriented plaquettes and all oriented $1 \times 2$ planar loops respectively.

The overall weight $\omega_{\text{C}^\star}$ is $1$ if no C$^\star$ boundary conditions are used. With C$^\star$ boundary conditions $\omega_{\text{C}^\star}=1/2$ corrects for the double counting introduced by summing over all plaquette and double--plaquette loops in the extended lattice instead of the physical lattice. The coefficients $c_{0,1}$ satisfy the relation $c_0 + 8c_1 = 1$. The Wilson action is obtained by choosing $c_0 = 1$, the L\"uscher--Weisz action is obtained by choosing $c_0 = \tfrac{5}{3}$, and the Iwasaki action is obtained by choosing $c_0 = 3.648$.
The parameter $g_0$ is the bare SU(3) gauge coupling, and $e_0$ is the bare U(1) gauge coupling in terms of which the bare fine--structure constant is given by
\begin{gather}
   \alpha_0 = \frac{e_0^2}{4\pi} \ .
\end{gather}
In the compact formulation of QED, all electric charges must be integer multiples of some elementary charge $q_\text{el}$ which is defined in units of the charge of the positron. As discussed in ref.~\cite{Lucini:2015hfa}, $q_\text{el}$ appears as an overall factor in the gauge action and essentially sets the normalization of the U(1) gauge field in the continuum limit. Even though in infinite volume $q_\text{el}=1/3$ would be an appropriate choice in order to simulate quarks, in finite volume with C$^\star$ boundary conditions one needs to choose $q_\text{el}=1/6$ in order to construct gauge--invariant interpolating operators for charged hadrons.

\subsection{Dirac operator}
\label{subsec:dirac}

The Dirac operator can be written as
\begin{gather}
   D = m_0 + D_\text{w} + \delta D_\text{sw} + \delta D_\text{b}
   \ .
\end{gather}
where $D_\text{w}$ is the (unimproved) Wilson--Dirac operator, $\delta D_\text{sw}$ is the Sheikholeslami–Wohlert (SW) term, and $\delta D_\text{b}$ is the time boundary $O(a)$-improvement term. For simplicity, periodic boundary conditions in the time direction will be assumed, which means $\delta D_\text{b}=0$

In presence of electromagnetism, the Dirac operator depends on the electric charge of the quark field. Let $q$ be the physical electric charge in units of $e$ (i.e. $q=2/3$ for the up quark, and $q=-1/3$ for the down quark). In the compact formulation of QED, all electric charges must be integer multiples of an elementary charge $q_\text{el}$, which appears as a parameter in the U(1) gauge action~\eqref{eq:action:U1}. It is useful to introduce the integer parameter
\begin{gather}
   \hat{q} = \frac{q}{q_\text{el}} \in \mathbb{Z} \ .
\end{gather}
The Wilson--Dirac operator can be written as
\begin{gather}
   D_\text{w}
   =
   \sum_{\mu=0}^3 \frac{1}{2} \left\{ \gamma_\mu ( \nabla_\mu + \nabla^*_\mu ) - \nabla^*_\mu \nabla_\mu \right\}
   \ ,
\end{gather}
where the covariant derivatives are defined as
\begin{gather}
   \nabla_\mu \psi(x) = U(x,\mu) z(x,\mu)^{\hat{q}} \psi(x+\hat{\mu}) - \psi(x)
   \ , \\
   \nabla^*_\mu \psi(x) = \psi(x) - U(x-\hat{\mu},\mu)^\dag z(x-\hat{\mu},\mu)^{-\hat{q}} \psi(x-\hat{\mu})
   \ .
\end{gather}
Notice that the integer parameter $\hat{q}$ appears in the hopping term of the Wilson--Dirac operator. The SW term is given by
\begin{gather}
   \delta D_\text{sw}
   =
   c^\text{SU(3)}_\text{sw} \sum_{\mu,\nu=0}^3 \frac{i}{4} \sigma_{\mu\nu} \widehat{F}_{\mu\nu}
   + q \, c^\text{U(1)}_\text{sw} \sum_{\mu,\nu=0}^3 \frac{i}{4} \sigma_{\mu\nu} \widehat{A}_{\mu\nu}
   \ .
\end{gather}
The SU(3) field tensor $\widehat{F}_{\mu\nu}(x)$ and the U(1) field tensor $\widehat{A}_{\mu\nu}(x)$ are constructed in terms of the clover plaquette. The explicit expression of the SU(3) field tensor used in \texttt{openQ*D} can be found in ref.~\cite{Luscher:1996sc}, while the U(1) field tensor is given here,
\begin{gather}
   \widehat{A}_{\mu\nu}(x) = \frac{i}{4 q_\text{el}} \Im\left\{
   z_{\mu\nu}(x)+ 
   z_{\mu\nu}(x-\hat{\mu})+
   z_{\mu\nu}(x-\hat{\nu})+
   z_{\mu\nu}(x-\hat{\mu}-\hat{\nu})
   \right\}
   \ , \\
   z_{\mu\nu}(x) = z(x,\mu) z(x+\hat{\mu},\nu) z(x+\hat{\nu},\mu)^\dag z(x,\nu)^\dag
   \ .
\end{gather}
The normalization is chosen in such a way that $ -i e_0 \hat A_{\mu\nu}(x) $ is the canonically--normalized field tensor in the naive continuum limit, and the discretized field tensors are defined to be anti--hermitian. In the \texttt{openQ*D} code, it is possible to choose the  values for all parameters of the Dirac operator independently for each flavour. In particular notice that, in presence of electromagnetism, the values of $c^\text{SU(3)}_\text{sw}$ and $c^\text{U(1)}_\text{sw}$ must depend on the electric charge in order to obtain $O(a)$ improvement.

\subsection{Pseudofermion action}
\label{subsec:rhmc}

The sum over the extended lattice introduces a double counting which can be corrected with an extra $1/2$ factor in the fermion action:
\begin{gather}
   S_\text{f} = \frac{1}{2} \sum_{x \in \Lambda_\text{ext}} \bar{\psi}(x) D \psi(x) = - \frac{1}{2} \sum_{x \in \Lambda_\text{ext}} \psi(x)^T C \mathcal{T} D \psi(x) \equiv -\frac{1}{2} \psi^T C\mathcal{T}D \psi \ ,
\end{gather}
where the orbifold constraint~\eqref{eq:majorana} has been used. By using the properties of the $C$ matrix one easily proves that
\begin{gather}
   D[U,z]^T = C D[U^*,z^*] C^{-1} \ .
\end{gather}
If the gauge field respects the orbifold constraint, and $\mathcal{T}$ is the translation operator defined in~\eqref{eq:translation}, one trivially gets
\begin{gather}
   \mathcal{T} D[U^*,z^*] \mathcal{T}^{-1} = D[U,z] \ .
\end{gather}
By combining the previous two equations, and by using the fact that $C$ is anti--symmetric while $\mathcal{T}$ is symmetric, one gets that the matrix $C\mathcal{T}D$ is anti--symmetric. The integration over the fermion field yields the Pfaffian of $C\mathcal{T}D$ up to an irrelevant overall factor which will be reabsorbed in the definition of the fermionic integration measure,
\begin{gather}
   \int [ \d \psi ]_{\Lambda_\text{ext}} e^{- S_f} = \int [ \d \psi ]_{\Lambda_\text{ext}} e^{\frac{1}{2} \psi^T C\mathcal{T}D \psi} = \text{Pf}\, (C\mathcal{T}D) \ .
\end{gather}

In the continuum limit, the Pfaffian of $C\mathcal{T}D$ is positive \cite{Lucini:2015hfa}. However at fixed lattice spacing, the Pfaffian is shown to be real but it can be negative on rough enough gauge configurations. The absolute value of the Pfaffian of $C\mathcal{T}D$ has a representation in terms of the determinant of a positive operator (notice that $\det C = \det \mathcal{T} = 1$)
\begin{gather}
   \left| \text{Pf}\, (C\mathcal{T}D) \right|
   =
   \left| \text{Det}\, (C\mathcal{T}D) \right|^{1/2}
   =
   \text{Det}\, (D^\dag D)^{1/4}
   \ .
\end{gather}

The \texttt{openQ*D-0.9a2} code uses even--odd preconditioning. After decomposing the lattice in even and odd sites in the standard way, the Dirac is represented in block form as
\begin{gather}
   D =
   \begin{pmatrix}
      D_\text{ee} & D_\text{eo} \\
      D_\text{oe} & D_\text{oo}
   \end{pmatrix}
   \ .
\end{gather}
In terms of the even--odd preconditioned Dirac operator
\begin{gather}
   \hat{D} = D_\text{ee} - D_\text{eo} D_\text{oo}^{-1} D_\text{oe} \ ,
\end{gather}
the absolute value of the Pfaffian can be written as
\begin{gather}
   \left| \text{Pf}\, (C\mathcal{T}D) \right|
   =
   \left| \text{Det}\, D \right|^{1/2}
   =
   \left| \text{Det}\, D_\text{oo} \ \text{Det}\, \hat{D} \right|^{1/2}
   =
   \left| \text{Det}\, D_\text{oo} \right|^{1/2} \ \text{Det}\, (\hat{D}^\dag \hat{D})^{1/4}
   \ .
\end{gather}

In order to stabilize the configuration generation, the \texttt{openQ*D-0.9a2} code allows to introduce a parameter $\hat{\mu}^2$ and to replace
\begin{gather}
   \text{Det}\, (\hat{D}^\dag \hat{D})^{1/4}
   \to
   \text{Det}\, (\hat{D}^\dag \hat{D} + \hat{\mu}^2)^{1/4}
   \ ,
\end{gather}
which can be corrected by introducing a properly--defined reweighting factor in the observables (following the strategy of ref.~\cite{Luscher:2012av}). Let $R$ be a rational approximation of order $[N,N]$ of
\begin{gather}
   R \simeq (\hat{D}^\dag \hat{D} + \hat{\mu}^2)^{-1/4} \ .
\end{gather}
The  \texttt{openQ*D-0.9a2} inherits from \texttt{openQCD-1.6} the frequency splitting for the RHMC~\cite{Luscher:2012av}. If the rational approximation is written explicitly as
\begin{gather}
   R = A \prod_{j=1}^N \frac{\hat{D}^\dag \hat{D} + \nu_j^2}{\hat{D}^\dag \hat{D} + \mu_j^2}
   \ ,
\end{gather}
where the finite sequences $\mu_j$ and $\nu_j$ are assumed to be monotonically increasing, one can always define a factorization
\begin{gather}
   R = A \, P_1 P_2 \cdots P_n \ ,
\end{gather}
where the factors $P_k$ contain all the zeroes and poles with indices in a given range $\{ J_k , J_k+1 , \dots , J_{k+1}-1 \}$, i.e.
\begin{gather}
   P_k = \prod_{j=J_k}^{J_{k+1}-1} \frac{\hat{D}^\dag \hat{D} + \nu_j^2}{\hat{D}^\dag \hat{D} + \mu_j^2} \ .
\end{gather}

The  \texttt{openQ*D-0.9a2} code simulates the absolute value of the Pfaffian by means of the pseudofermion action given by
\begin{gather}
   \left| \text{Pf}\, (CTD) \right|
   \simeq 
   \left| \text{Det}\, D_\text{oo} \right|^{1/2}
   \ 
   \text{Det}\, R^{-1}
   =
   \int [\d \phi \, \d \phi^*]_{\text{even}(\Lambda_\text{ext})} \ e^{-S_\text{pf}}
   \ , \\
   S_\text{pf}
   =
   - \frac{1}{2} \ln \left| \text{Det}\, D_\text{oo} \right|
   + \sum_{k=1}^n \phi_k^\dag P_k \phi_k
   \ .
\end{gather}
The pseudofermion $\phi_k$ is a complex field with gauge and spinor indices, which lives on the even sites of the extended lattice. It satisfies periodic boundary conditions in direction 1, and possibly shifted boundary condition in directions $k=2,3$. It is completely unrestricted over the extended lattice. Notice that, since $D_\text{oo}$ is diagonal in the lattice index, its determinant can be calculated exactly.

\subsection{Molecular dynamics}
\label{subsec:md}

The momentum fields associated to the SU(3) and U(1) gauge field are denoted by $\Pi(x,\mu)$ and $\pi(x,\mu)$ respectively. The momentum $\Pi(x,\mu)$ lives in the Lie algebra of SU(3),
\begin{gather}
   \Pi(x,\mu) = \Pi^a(x,\mu) T^a \ ,
\end{gather}
where $\Pi^a(x,\mu)$ are taken to be real. The momentum field $\pi(x,\mu)$ is taken to be real, like the gauge field $A(x,\mu)$. With C$^\star$ boundary conditions, the momentum fields must satisfy the orbifold constraint
\begin{gather}
   \label{eq:orbi:mom}
   \Pi(x + L_1 \hat{e}_1,\mu) = \Pi(x,\mu)^* \ ,
   \qquad
   \pi(x + L_1 \hat{e}_1,\mu) = -\pi(x,\mu) \ .
\end{gather}
Summing over all momenta in the extended lattice instead of the physical lattice introduces a double counting which can be corrected with an extra $1/2$ factor in the MD Hamiltonian:
\begin{gather}
   H = \frac{1}{4} \sum_{x,\mu} \bigg\{ [\pi(x,\mu)]^2 + \sum_a [\Pi^a(x,\mu)]^2 \bigg\} + S(U,A) \ .
\end{gather}

In deriving the MD equations, one needs to take into account the orbifold constraint. It is convenient to define the forces as the derivative of the action $S(U,A)$ thought as a function of the unconstrained fields on the extended lattice
\begin{gather}
   \label{eq:orbi:force}
   F(x,\mu) = - \left[ \partial_{U(x,\mu)} S(U,A) \right]_\text{unconstrained} \ ,
   \qquad
   f(x,\mu) = - \left[ \partial_{A(x,\mu)} S(U,A) \right]_\text{unconstrained} \ .
\end{gather}
By using the orbifold constraint and the chain rule, the MD equations are found to be
\begin{subequations}
   \label{eq:orbi:MD}
   \begin{alignat}{2}
      &
      \partial_t U(x,\mu) = \Pi(x,\mu) U(x,\mu) \ ,
      &&
      \partial_t \Pi(x,\mu) = F(x,\mu) + F(x + L_1 \hat{e}_1,\mu)^* \ ,
      \\
      & 
      \partial_t A(x,\mu) = \pi(x,\mu) \ ,
      & \qquad &
      \partial_t \pi(x,\mu) = f(x,\mu) - f(x + L_1 \hat{e}_1,\mu) \ .
   \end{alignat}
\end{subequations}
Since the Hamiltonian is invariant under translations and charge--conjugation, the orbifold constraint is preserved by the MD. In fact the orbifold constraint is also preserved by the discrete integrators used by the \texttt{openQ*D} code.

\section{QCD: some tests}
\label{sec:qcd}

We have performed some test runs in order to assess two issues:
\begin{itemize}
   \item the overhead due to the orbifold construction;
   \item the effectiveness of deflation with C$^\star$ boundary conditions.
\end{itemize}
The results presented here allow only to scratch the surface of these issues. The drawn conclusion are far from definitive and should be taken with a grain of salt.

All runs presented in this section share the following parameters:
\begin{itemize}
   \item QCD with $N_f=2$ degenerate flavours;
   \item Physical lattice $64 \times 32^3$ with periodic boundary conditions in time;
   \item Wilson action with $\beta=5.2$;
   \item Non--perturbatively $O(a)$ improved Wilson fermions with $\kappa=0.1359$ and $c_\text{sw}=2.017147$;
   \item MD trajectory lenght $\tau=2$;
   \item $a \simeq 0.08\text{ fm}$, $m_\pi \simeq 380\text{ MeV}$, $m_\pi L \simeq 4.7$.
\end{itemize}
The parameters and the estimates for the lattice spacing and the pion mass are taken from the A4 ensemble in ref.~\cite{Fritzsch:2012wq} (see table 2). The $c_\text{sw}$ improvement coefficient is calculated by using eq. (2.25) in ref.~\cite{Jansen:1998mx}. In all cases we have used a twisted--mass reweighting with $\hat{\mu}=0.001$.

\subsection{Overhead due to the orbifold construction}

A possible and direct way to implement C$^\star$ boundary conditions is to pack the quark and antiquark into a doublet
\begin{gather}
   \Psi =
   \begin{pmatrix}
      \psi \\
      C^{-1} \bar{\psi}^T
   \end{pmatrix}
   \ ,
\end{gather}
which transform under the $\Box \times \bar{\Box}$ representation of the gauge group. The boundary conditions swap the two components of $\Psi$. When the pseudofermion action is derived, pseudofermions will have two components as well. Let us denote by $D_\text{C,2c}$ the Dirac operator with C$^\star$ boundary conditions in the 2-component formulation. The Dirac operator acts independently on the two components in the bulk and mixes the two components at the boundary.

In \texttt{openQ*D-0.9a2}, C$^\star$ boundary conditions are implemented through an orbifold construction. The Dirac operator $D_\text{C,orbi}$ acts on single--component pseudofermions which are defined on a lattice that is twice as large as the physical lattice.

The orbifold construction was preferred to the two--component formulation for the \texttt{openQ*D-0.9a2} code, since it required no modification of the Dirac operator and solvers of the original \texttt{openQCD-1.6} code. The implementation of C$^\star$ boundary conditions via the orbifold construction is almost trivial. However, one may think that simulations with the orbifold construction are much more expensive. The question we want to address is: \textbf{how more expensive is the orbifold construction with respect to the two--component construction?}

If the physical volume $V=TL^3$ is kept constant it is obvious that
\begin{gather}
   \text{cost} [ D_\text{C,2c} \Phi_\text{2c,V} ] = \text{cost} [ D_\text{C,orbi} \phi_\text{2c,2V} ] = 2 \ \text{cost} [ D_\text{P} \phi_\text{1c,V} ] \ \{ 1 + O(L^{-1}) \} \ ,
\end{gather}
where $D_\text{P}$ is the standard operator with periodic boundary conditions, which acts on single--component pseudofermions. Therefore, the application of the Dirac operator and the solution of the Dirac equation has exactly the same cost in the two possible implementations of C$^\star$ boundary conditions.

The orbifold construction loses for two reasons:
\begin{itemize}
   \item The gauge fields are evolved twice, in the physical and mirror lattice, see fig.~\ref{fig:geometry}. This operation is relatively cheap, but it is done many times (at the innermost level of the integrator).
   \item The forces in the physical and mirror lattices need to be summed in the evolution equations for the momenta, see eqs.~\eqref{eq:orbi:MD}, and this requires MPI communications. The \texttt{openQ*D-0.9a2} code implements two solutions to mitigate this issue. First we notice that the gauge force (which is integrated more often) satisfies automatically the orbifold constraint. The sum of forces in eqs.~\eqref{eq:orbi:MD} gives a trivial factor of two for the gauge force. As a second measure, the MPI ranks are organized in such a way that a point $x$ and its mirror point $x + L_1\hat{e}_1$ belong to different MPI processes, but end up on the same multi--core node (assuming an MPI implementation for which MPI processes residing on a node are numbered consecutively). In this case the MPI communications needed for eqs.~\eqref{eq:orbi:MD} should not go through the network.
\end{itemize}
The cost for the gauge field and momenta evolution is identical in the 2-component implementation of the C$^\star$ boundary conditions and in the periodic case. Assuming that the volume is large enough, the cost of the simulation of the theory with periodic boundary conditions is going to be equal to the cost of the simulation of the theory with C$^\star$ boundary conditions implemented with the 2-component formalism, provided that all parameters of the algorithm are identical (in particular the RHMC algorithm is used in both cases, and the rational approximation is the same).

Then the overhead due to the orbifold construction is obtained by comparing the run with periodic boundary conditions (QCD2) with the run with C$^\star$ boundary conditions implemented with the orbifold construction (QCD3). For details about these runs, refer to app.~\ref{app:QCD}. By comparing the simulation times listed in table~\ref{tbl:qcdruns}, we conclude that the orbifold construction produces a relative overhead in cost of about $1\%$. The relative overhead is expected to decrease at smaller quark masses (since the solution of the Dirac equation becomes more expensive relatively to the gauge field and momenta evolution).

\subsection{RHMC vs. HMC}

In the case of $N_f=2$ degenerate flavours and periodic boundary conditions in space, one can use the HMC algorithm. We have generated the QCD1 ensemble with the HMC and twisted--mass Hasenbusch preconditioning, with splitting of the various terms in different integration levels.

In case of C$^\star$ boundary conditions one is forced to use the RHMC since the Pfaffian needs to be simulated. We use a representation of the Pfaffian of the type
\begin{gather}
   |\text{Pf}\, (C\mathcal{T}D)|^2 = \frac{1}{\text{Det}\, (D^\dag D)^{-1/4}} \frac{1}{\text{Det}\, (D^\dag D)^{-1/4}} \ ,
\end{gather}
with a rational approximation for $(D^\dag D)^{-1/4}$, as this is more similar to the setup needed for the inclusion of isospin--breaking corrections. We have generated the QCD3 ensemble with C$^\star$ boundary conditions, the RHMC with a rational approximation for $(D^\dag D)^{-1/4}$ and frequency splitting in different integration levels.

For details about the QCD1 and QCD3 runs, refer to app.~\ref{app:QCD}. By comparing the simulation times listed in table~\ref{tbl:qcdruns}, we see that the QCD3 run is about 5.5 times slower than the QCD1 run. This big factor is most likely an indication of the fact that the QCD3 run has not been optimized as well as the QCD1 run. We plan to invest more time into the optimization of the RHMC runs, and in particular we thank Kate Clark for providing useful suggestions during the conference.

\section{QCD+QED: some tests}
\label{sec:qed}

We have also used \texttt{openQ*D-0.9a2} code to perform test runs that include the dynamical degrees of freedom of the U(1) gauge field. The aim of these runs was twofold:
\begin{itemize}
\item  to examine whether any unexpected features appear if compact QCD+QED simulations
are performed with Wilson fermions and C$^\star$  boundary conditions;
\item  to get a first insight into autocorrelations of SU(3) and U(1) gauge observables.
\end{itemize}

\begin{figure}[tp]
\centering
 {\includegraphics[width=0.475\textwidth,clip]{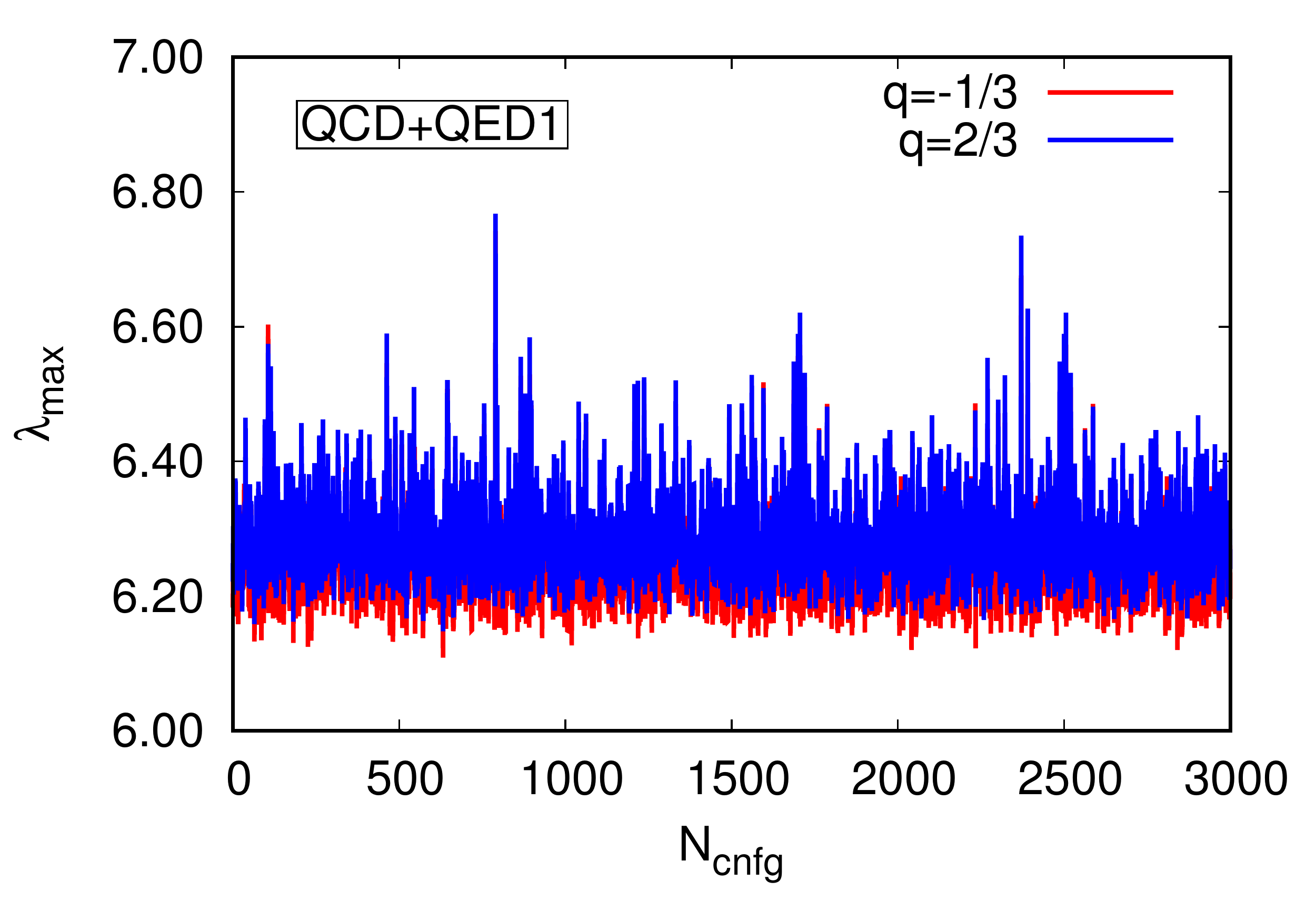}}\hfill
 {\includegraphics[width=0.475\textwidth,clip]{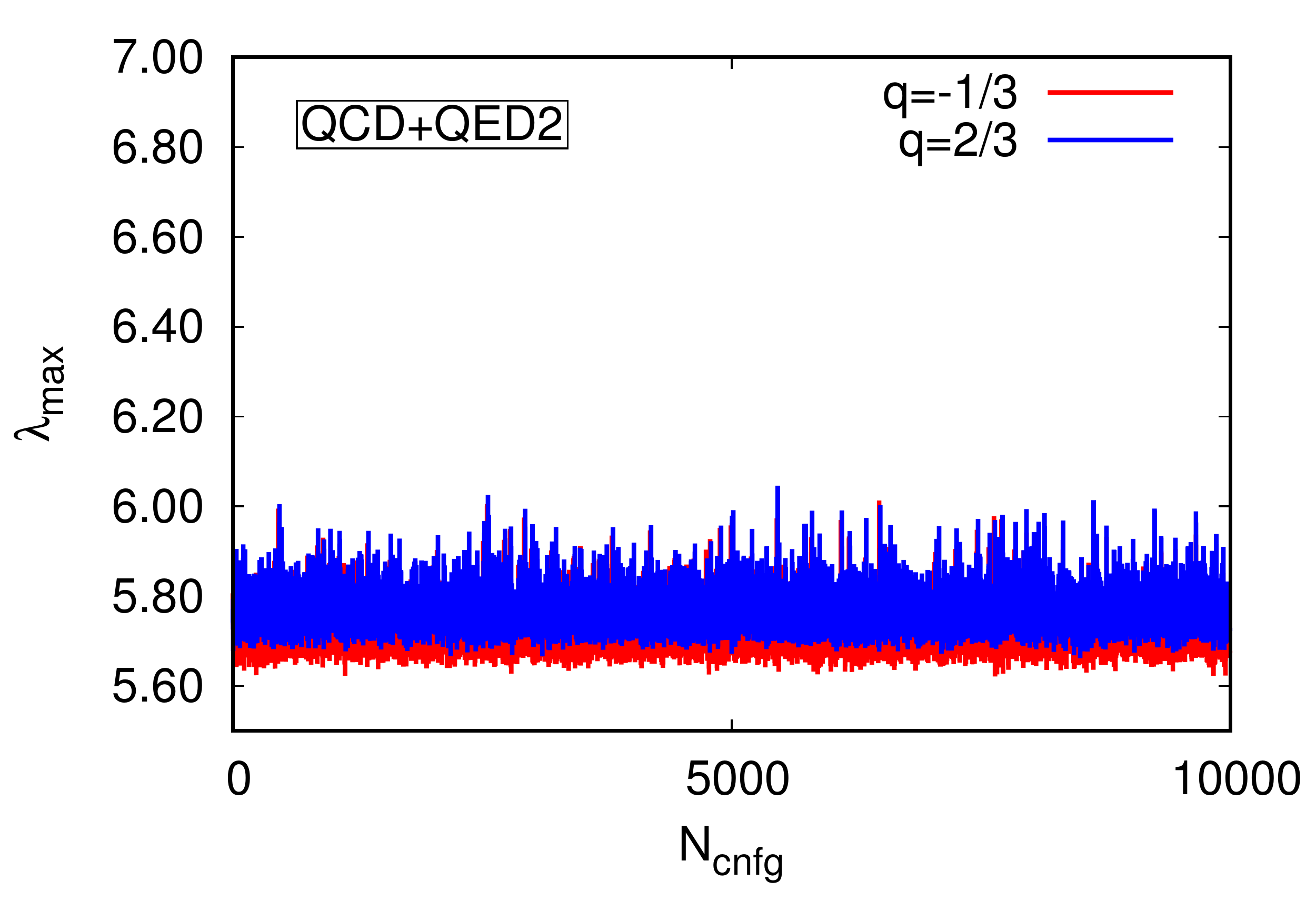}} ~\\[-3mm] 
 {\includegraphics[width=0.475\textwidth,clip]{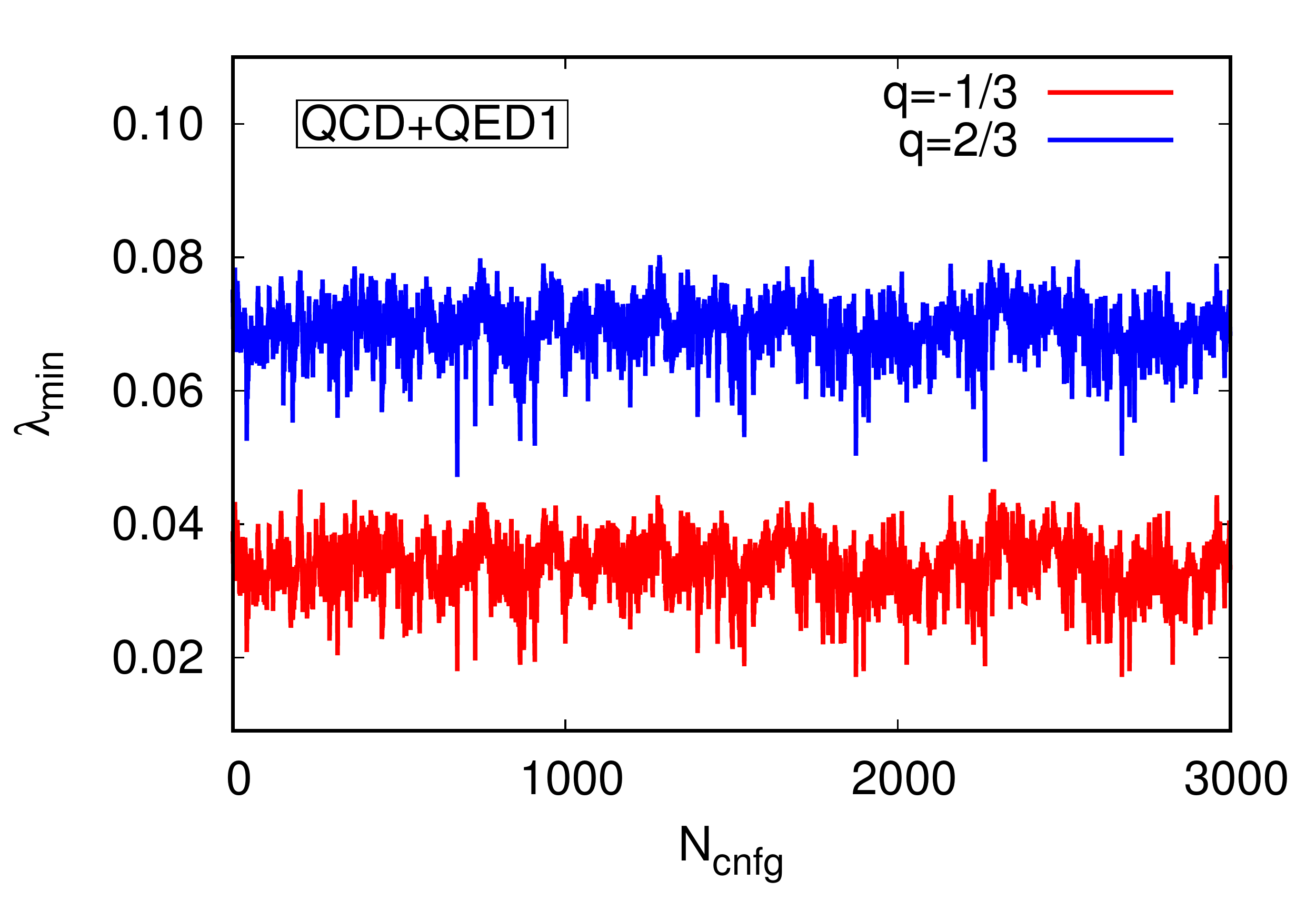}}\hfill
 {\includegraphics[width=0.475\textwidth,clip]{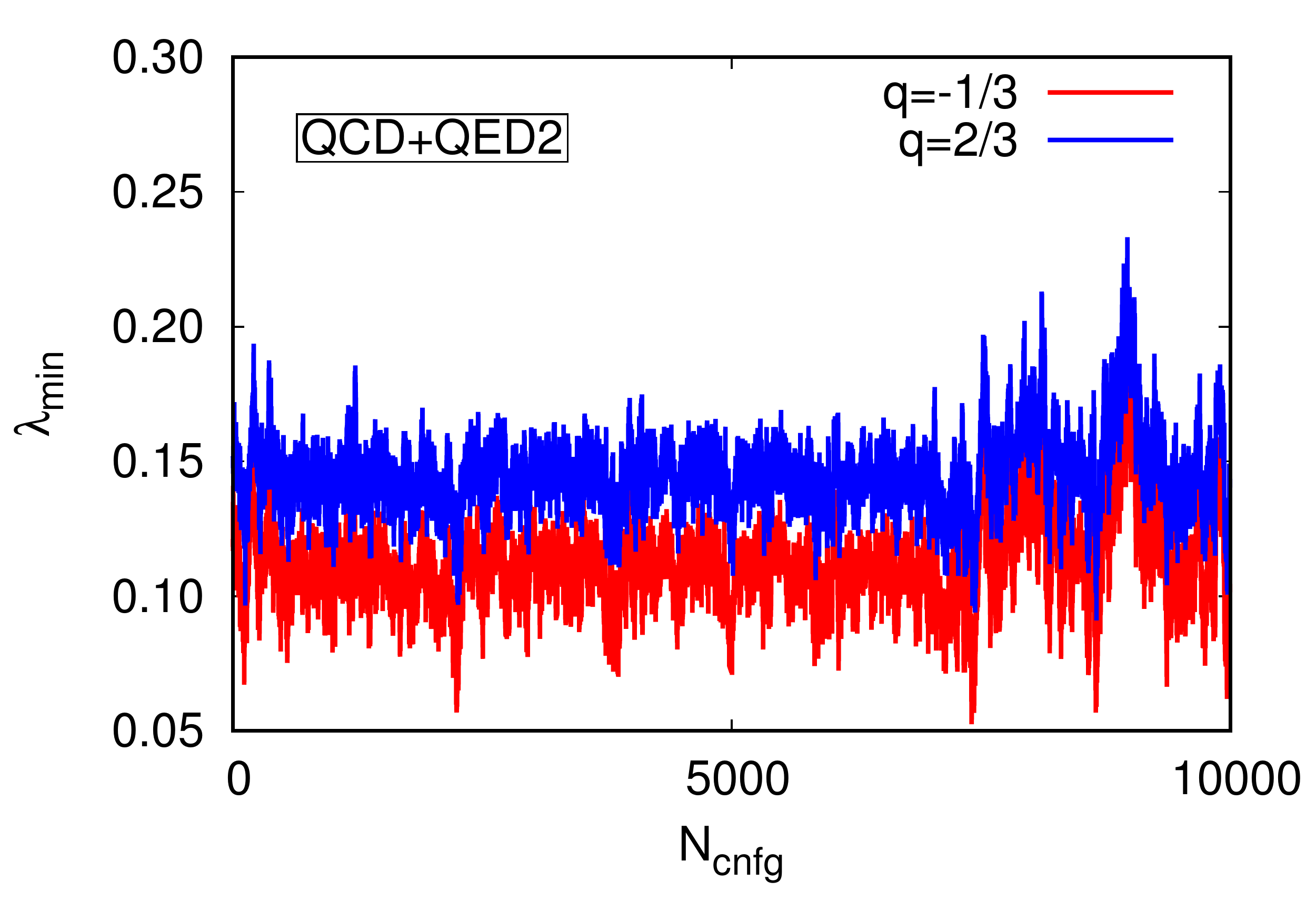}}\hfill
\vspace*{-4mm}
\caption{\small\textbf{Spectral ranges of $|\gamma_5 D|$ in the performed QCD+QED simulations.} The \textit{upper panels} represent the estimate of the highest eigenvalue of $|\gamma_5 D|$ corresponding to the $d/s$ quarks (red) and $u$ quark (blue). Similarly, the \textit{lower panels} represent the evolution of the estimated smallest eigenvalues.
\label{fig:spectral}
}
\vspace*{-4mm}
\end{figure}

The \texttt{openQ*D-0.9a2} code currently implements only the compact formulation of QED; hence, we perform the initial tests with the compact QED action. 
The test runs presented in this section share the following parameters:
\begin{itemize}
   \item QCD+QED with $N_f=2+1$ fermion flavours; 
  \item C$^\star$ bc’s in all space directions;
\item $\alpha_\text{em}=0.05 \approx 7 \alpha_\text{em}^\text{phys}$;
   \item L\"uscher--Weisz SU(3) gauge action and Wilson U(1) gauge action;
\item Wilson fermion action with SU(3) SW--term coefficients determined at $\alpha_\text{em}=0$;
   \item Tree--level coefficients for the U(1) SW--term;  
   \item MD trajectory length $\tau=0.7071$ 
(to be able to compare with \texttt{HiRep} code \cite{hansen:lat2017}).
\end{itemize}

The first run (\Qone) takes over the parameters from the H200 ensemble of the $N_f=2+1$ CLS  \cite{Bruno:2014jqa}, except that the lattice extent is halved in each of the space--time directions.  
The dynamical U(1) degrees of freedom contribute to the renormalization of the bare parameters, hence the estimate for the lattice spacing and pion mass cannot be taken from the CLS ensembles\footnote{Had the U(1) d.o.f. been switched off ($\alpha_\text{em} = 0$), the chosen parameter set would correspond to $m_\pi \approx 420 \text{ MeV}$.}, but rather need to be estimated independently. 

The parameters characteristic for each of the performed QCD+QED runs are the following:
\begin{itemize}
 \item \Qone: Physical lattice $32 \times 16^3$ with periodic bc's in time, $\beta=3.55$, $c_\text{sw,u}^\text{SU(3)}=c_\text{sw,d}^\text{SU(3)}=c_\text{sw,s}^\text{SU(3)}=1.824865$, $\kappa_\text{u}=\kappa_\text{d}=\kappa_\text{s}=0.137$;
 \item \Sone: Physical lattice $16 \times 8^3$ with open--SF bc's in time; $\beta=4.0$,  $c_\text{sw,u}^\text{SU(3)}=c_\text{sw,d}^\text{SU(3)}=c_\text{sw,s}^\text{SU(3)}=1.540714371185832$, $\kappa_\text{u}=\kappa_\text{d}=\kappa_\text{s}=0.136646552997824$;.
\end{itemize}
Although \texttt{openQ*D-0.9a2} code allows for twisted--mass reweighting, the runs described in this section do not use that option ($\hat{\mu}=0.0$). 
In both runs all three bare sea quark masses are taken to be the same. However, due to the differences in quark charges we end up with a degenerate pair of quarks (down and strange) with $q=-1/3$, and a single quark  (up) with $q=2/3$ in our simulations; hence, we are essentially simulating $N_f=2+1$ theory. 

\subsection{Spectral ranges of the Dirac operators}

We monitored the smallest and the largest eigenvalue of $|\gamma_5 D_\text{u}|$ and  $|\gamma_5 D_\text{d/s}|$ throughout the performed QCD+QED runs, in order to confirm that the spectral ranges of the rational approximations have been chosen correctly. It turns out that, subsequent to the thermalization phase, the inclusion of electromagnetic effects does not lead to large fluctuations of the spectral range. The results of the estimated spectral ranges in runs \Qone~ (left panels) and \Sone~(right panels) are shown in fig. \ref{fig:spectral}. See app. \ref{app:QCD-QED} for details on the chosen spectral ranges in the corresponding rational approximations. 

\begin{figure}[tp]
\centering
 {\includegraphics[width=0.310\textwidth,clip]{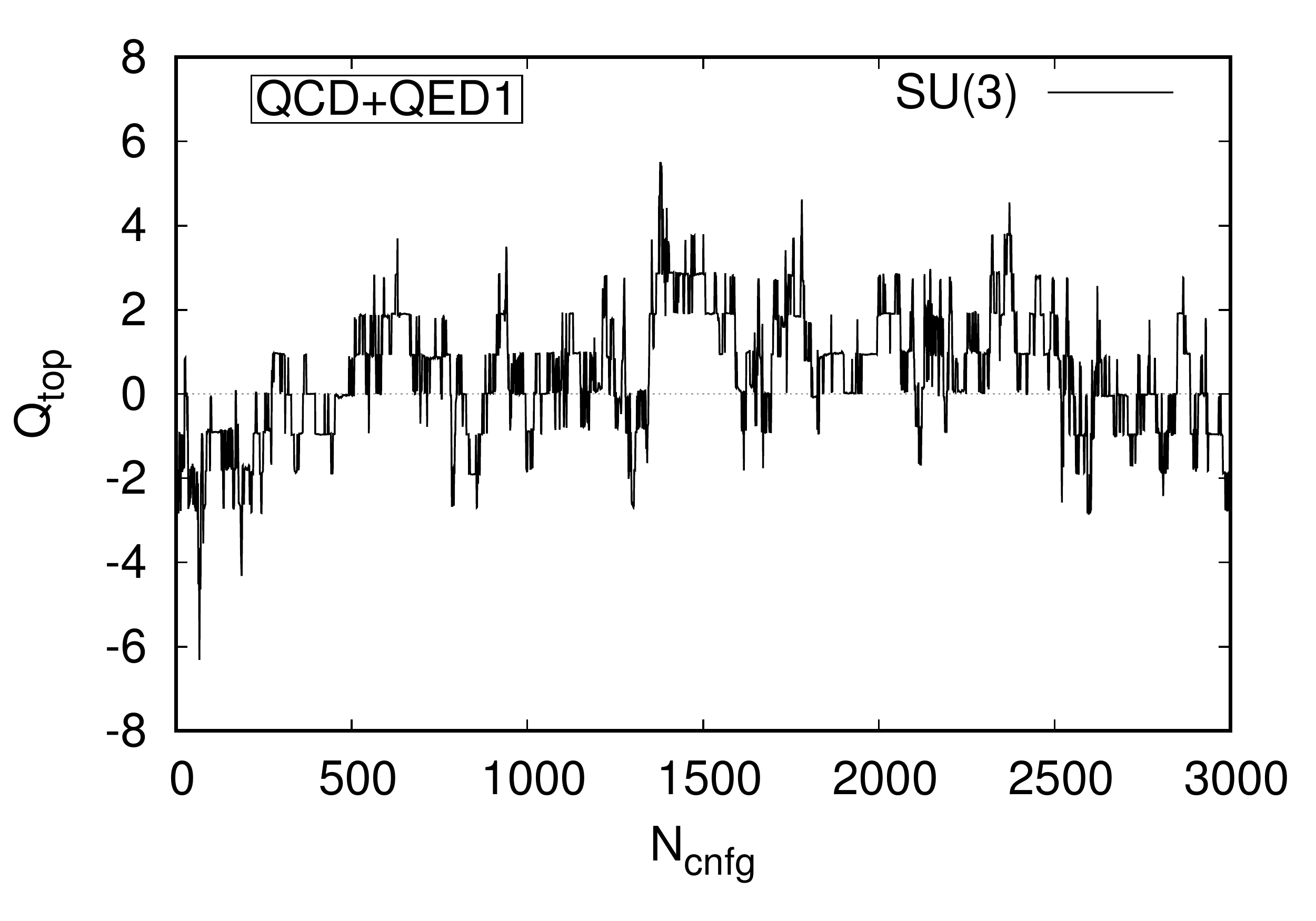}}\hfill
 {\includegraphics[width=0.310\textwidth,clip]{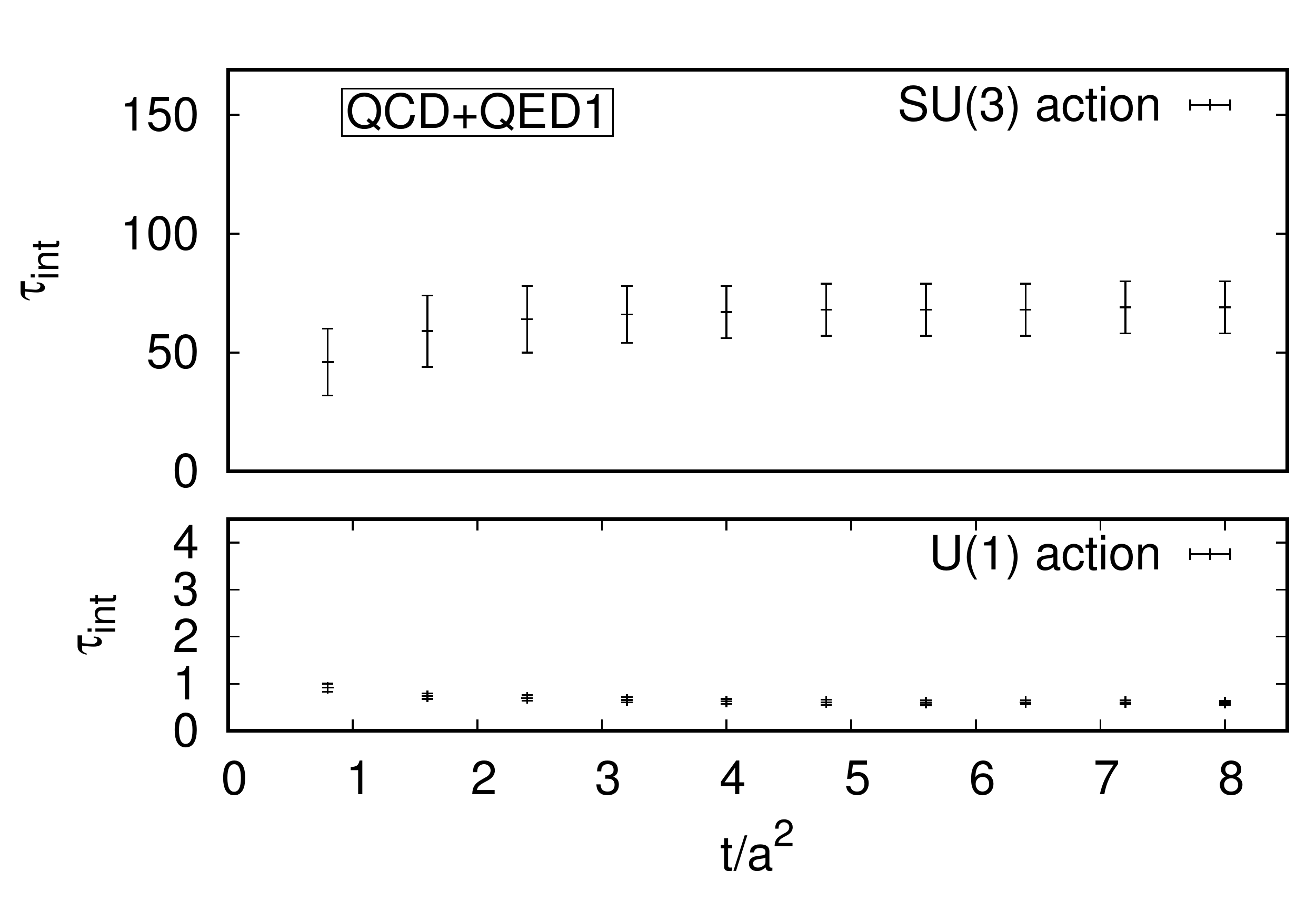}}\hfill
 {\includegraphics[width=0.310\textwidth,clip]{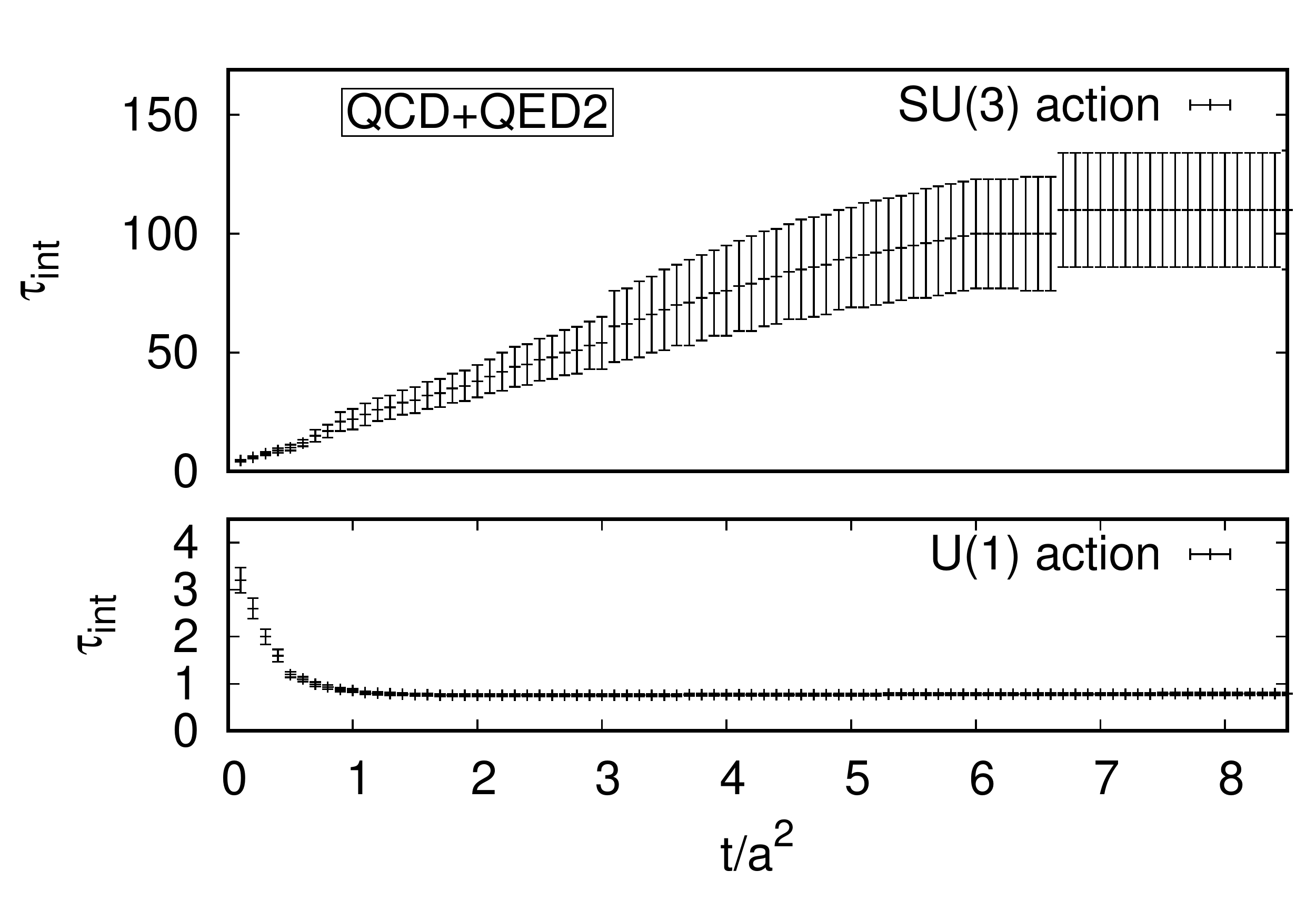}}\hfill
\vspace*{-4mm}
\caption{\small\textbf{Autocorrelations in the performed QCD+QED simulations.} The \textit{left panel} shows  the history of the SU(3) topological charge defined in eq. \eqref{eq:qtop}, measured on thermalized configurations in the \Qone~ensemble at the flow time $t$ defined by $\sqrt{8t}=0.3\times L$, and corresponding to $t/a^2=0.36$. The remaining plots compare the autocorrelations of the SU(3) and U(1) observables in the \Qone~run with periodic bc's in time (\textit{middle panel}) and in the small volume \Sone~run with open--SF bc's in time (\textit{right panel}). All runs feature C$^\star$  spatial bc's.
\label{fig:qtop}
}
\vspace*{-4mm}
\end{figure}

\subsection{Autocorrelations of SU(3) and U(1) gauge observables}

We use Wilson flow to define the global topological charge and  the action density for SU(3) and U(1) gauge fields (see app. \ref{app:QCD-QED} and ref. \cite{Luscher:2010iy}).
In the left panel of fig. \ref{fig:qtop} we show the history of $Q_\text{top}(t)$  in the ensemble \Qone, while the flow time dependence of autocorrelation times of the U(1) and SU(3) action in the same run are shown in the middle panel of fig. \ref{fig:qtop}. The flow time $t$ in the plotted $Q_\text{top}(t)$ is chosen such that $\sqrt{8t}=0.3\times L$. 
The \Qone~ run features periodic boundary conditions in time and the length of the run seems not to be sufficient to give a reliable estimate of the  autocorrelations of the global topological charge. 
This motivated the choice of input parameters for our second ensemble: open--SF bc's in time, and much smaller volume in physical units (smaller lattice $8^3\times16$ and larger value of $\beta$). 
The statistics gathered  in the \Sone~ run O(10000) allows for a good estimate of how fast the Wilson flow quantities decorrelate in the QCD+QED simulations with C$^\star$  space bc's. 
The autocorrelations of the U(1) and SU(3) action, see eqs. \eqref{eq:u1flow} and   \eqref{eq:su3flow}, for the run \Sone~ are shown in the right panel of fig. \ref{fig:qtop}. 
Notice that for both choices of bc's in time, the autocorrelations of U(1) observables are of $O(1)$ even though we are not using any technique to decrease the U(1) field autocorrelations, such as Fourier acceleration.

~\\[-2mm]
{\bf Acknowledgements.}
Most simulations reported in this paper were performed on a dedicated PC cluster at CERN. We are grateful to the CERN management for funding this machine and to the CERN IT Department for technical support. We also acknowledge Santander Supercomputacion support group at the University of Cantabria for providing access to Altamira Supercomputer at the Institute of Physics of Cantabria (IFCA-CSIC), member of the Spanish Supercomputing Network, for performing simulations/analyses. This work was supported by a grant from the Swiss National Supercomputing Centre (CSCS) under project ID s642 and by TCHPC (Research IT, Trinity College Dublin). Some calculations were performed on the Lonsdale cluster maintained by the Trinity Centre for High Performance Computing. This cluster was funded through grants from Science Foundation Ireland. I.C. thanks the EC for funding via the H2020 project INDIGO-Datacloud (RIA 653549).

\appendix

\section{Some details on the QCD test runs}
\label{app:QCD}

We have produced three ensembles which differ by the space boundary conditions (and consequently by the simulated lattice) and by the algorithm used, as summarized in table~\ref{tbl:qcdruns}. In all cases, a 3--level integrator has been used, following the strategy used in ref.~\cite{Bruno:2014jqa}:
\begin{itemize}
   \item 1 steps of 4th order OMF integrator in the innermost level (\textit{level 0});
   \item 1 step of 4th order OMF integrator in the intermediate level (\textit{level 1});
   \item 12 step of 2nd order OMF integrator with $\lambda=1/6$ in the outermost level (\textit{level 2}).
\end{itemize}
Only the gauge force is integrated in level 0, while the fermionic forces are distributed in levels 1 and 2, in different ways depending on the particular algorithm used. In all cases the acceptance rate is found to be between $90\%$ and $95\%$. Let us look now in detail at the algorithms used for each run listed in table~\ref{tbl:qcdruns}.

\begin{table}[htb]
  \caption{$N_f=2$ QCD test runs ($a \simeq 0.08\text{ fm}$, $m_\pi \simeq 380\text{ MeV}$, $m_\pi L \simeq 4.7$). Notice that, in case of C$^\star$ boundary conditions, the global lattice is larger than physical lattice because of the orbifold contruction.}
  \label{tbl:qcdruns}
  \centering
  \begin{small}
  \begin{tabular}{llllll}\toprule
  Run name & Space bc's & Global lattice & Local lattice & Algorithm & Time/trajs \\\midrule
  QCD1 & periodic & $64 \times 32^3$ & $8 \times 8^3$ & HMC$+$TM & 600s \\
  QCD2 & periodic & $64 \times 32^3$ & $8 \times 8^3$ & RHMC--4$\times$1/4 & 3890s \\
  QCD3 & C$^\star$ in 3 dirs & $64 \times 64 \times 32^2$ & $8 \times 16 \times 8^2$ & RHMC--2$\times$1/4 & 3920s \\\bottomrule
  \end{tabular}
  \end{small}
\end{table}

\textbf{QCD1}. In case of periodic boundary conditions, one can use the HMC algorithm with even--odd and Hasenbusch twisted--mass preconditioning. We use a pseudofermion action of the form
\begin{gather}
   S_\text{pf} = S_\text{pf,1} + S_\text{pf,2}
   \ , \\
   S_\text{pf,1} =
   - 2 \ln | \det D_\text{oo} |
   + \phi_3^\dag \frac{1}{\hat{D}^\dag \hat{D} + \mu_3} \phi_3
   + \phi_2^\dag \frac{\hat{D}^\dag \hat{D} + \mu_3}{\hat{D}^\dag \hat{D} + \mu_2} \phi_2
   + \phi_1^\dag \frac{\hat{D}^\dag \hat{D} + \mu_2}{\hat{D}^\dag \hat{D} + \mu_1} \phi_1
   \ , \\
   S_\text{pf,2} = \phi_0^\dag \frac{\hat{D}^\dag \hat{D} + \mu_1}{\hat{D}^\dag \hat{D} + \hat{\mu}} \phi_0
   \ .
\end{gather}
We choose $(\mu_1,\mu_2,\mu_3) = (0.005,0.05,0.5)$. The force associated to $S_\text{pf,1}$ is integrated in level 1, while the force associated to $S_\text{pf,2}$ is integrated in level 2. The Dirac equation is solved by means of a conjugate gradient for twisted mass $\mu=0.5$ and by means of the deflation--accelerated solver in all other cases.
   
\textbf{QCD2}. In case of periodic boundary conditions, one can use the RHMC algorithm with multiple copies of pseudofermion fields. In particular, we construct the optimal rational approximation with relative precision of $10^{-6}$ for
\begin{gather}
   (\hat{D}^\dag \hat{D} + \hat{\mu}^2)^{-1/4} \ ,
\end{gather}
assuming that the spectrum of $|\gamma_5 \hat{D}|$ is included in the range $[1.98 \times 10^{-3}, 7.62]$. We split the rational approximation in 8 factors, as explained in subsection~\ref{subsec:rhmc}. The forces associated to different factors are integrated in different levels. Also different solvers are used to solve the Dirac equations associated to the different factors. These details are summarized in table~\ref{tbl:rational}. The pseudofermion action reads
\begin{gather}
   S_\text{pf}
   =
   - \frac{N_\text{pf}}{2} \ln \left| \text{Det}\, D_\text{oo} \right|
   + \sum_{\alpha=1}^{N_\text{pf}} \sum_{k=1}^8 \phi_{\alpha,k}^\dag P_k \phi_{\alpha,k}
   \ ,
\end{gather}
where $N_\text{pf}=4$ copies of pseudofermion fields have been used in order to reproduce the correct number of flavours.

\textbf{QCD3}. In case of C$^\star$ boundary conditions, one must use the RHMC algorithm with multiple copies of pseudofermion fields. We use the same setup as for the QCD2 run (see table~\ref{tbl:rational}) with the only difference that in this case the correct number of flavours is obtained by using $N_\text{pf}=2$ copies of pseudofermion fields.

\begin{table}[t]
  \centering
  \caption{Rational approximation used for the QCD2 and QCD3 runs. Each zero/pole pair is included in a frequency--splitting (FS) factor $P_k$, the corresponding pseudofermion force is integrated in either level 1 or level 2, and different solvers are used to solve the associated Dirac equations.}
  \label{tbl:rational}
\begin{small}
  \begin{tabular}{llllll}\toprule
  $j$ & $\nu_j$ & $\mu_j$ & FS factor & Integration level & Solver \\\midrule
  1  & $1.7626 \times 10^{+01}$ & $1.2383 \times 10^{+01}$ & $P_1$ & level 1 &   \rdelim\}{7}{3mm}[Multi--shift CG] \\
  2  & $5.9094 \times 10^{+00}$ & $4.8370 \times 10^{+00}$ & $P_1$ & level 1 & \\
  3  & $2.7961 \times 10^{+00}$ & $2.3515 \times 10^{+00}$ & $P_1$ & level 1 & \\
  4  & $1.4167 \times 10^{+00}$ & $1.1994 \times 10^{+00}$ & $P_1$ & level 1 & \\
  5  & $7.3020 \times 10^{-01}$ & $6.1927 \times 10^{-01}$ & $P_1$ & level 1 & \\
  6  & $3.7808 \times 10^{-01}$ & $3.2079 \times 10^{-01}$ & $P_1$ & level 1 & \\
  7  & $1.9600 \times 10^{-01}$ & $1.6632 \times 10^{-01}$ & $P_1$ & level 1 & \\
  8  & $1.0163 \times 10^{-01}$ & $8.6246 \times 10^{-02}$ & $P_2$ & level 1 & Deflation--accelerated \\
  9  & $5.2700 \times 10^{-02}$ & $4.4718 \times 10^{-02}$ & $P_3$ & level 2 & Deflation--accelerated \\
  10 & $2.7313 \times 10^{-02}$ & $2.3170 \times 10^{-02}$ & $P_4$ & level 2 & Deflation--accelerated \\
  11 & $1.4128 \times 10^{-02}$ & $1.1973 \times 10^{-02}$ & $P_5$ & level 2 & Deflation--accelerated \\
  12 & $7.2571 \times 10^{-03}$ & $6.1273 \times 10^{-03}$ & $P_6$ & level 2 & Deflation--accelerated \\
  13 & $3.6347 \times 10^{-03}$ & $3.0301 \times 10^{-03}$ & $P_7$ & level 2 & Deflation--accelerated \\
  14 & $1.6921 \times 10^{-03}$ & $1.3855 \times 10^{-03}$ & $P_8$ & level 2 & Deflation--accelerated \\
  \bottomrule
  \end{tabular}
  \end{small}
\end{table}

\section{Some details on the QCD+QED test runs}
\label{app:QCD-QED}

\begin{table}[b]
  \centering
  \caption{$N_f=2+1$ QCD+QED test runs. N represents the order of the corresponding rational approximation, and  $[\lambda_\text{min},\lambda_\text{max}]$ is the range that is assumed to include the spectrum of $|\gamma_5 \hat{D}|$ in both cases. 
}
  \label{tbl:qcdqedruns}  
\begin{small}
  \begin{tabular}{llllllll}\toprule
  Run name & Time bc's & Lattice & \multicolumn{2}{l}{$(D^\dag D)^{-1/2}$} & \multicolumn{2}{l}{$(D^\dag D)^{-1/4}$} &$N_\text{cnfg}$ \\\midrule
 $ ~$  &  $ ~$ & $ ~$&N&$[\lambda_\text{min},\lambda_\text{max}]$&N&$[\lambda_\text{min},\lambda_\text{max}]$ &  ~\\\midrule
  QCD+QED1&periodic & $32 \times 16^3$ & $ 20 $ & $[9.17\times 10^{-3}, 6.94]$ & 16& $[3.65\times 10^{-2},6.94]$ &  $ 3000$ \\
  QCD+QED2&open-SF & $16 \times 8^3$ & $ 20 $ & $[9.17\times 10^{-3}, 6.94]$  & 16 & $[3.65\times 10^{-2},6.94]$ &  $ 15000$ 
\\\bottomrule
  \end{tabular}
\end{small}
\end{table}

We have produced two 
ensembles which differ by the time boundary conditions and by the lattice size. The algorithm used in all cases is the same (RHMC) and involves a 3--level integrator:
\begin{itemize}
   \item 1 step of 4th order OMF integrator in the innermost level (\textit{level 0});
   \item 3 steps of 4th order OMF integrator in the intermediate level (\textit{level 1});
   \item 10 steps of 4th order OMF integrator in the outermost level (\textit{level 2}).
\end{itemize}
The U(1) force is integrated in level 0, 
the SU(3) force in level 1, and all the fermionic forces are integrated in the outermost level (level 2). 
The two runs are simulating C$^\star$ bc's in space and the Dirac equation is solved by means of CG in all cases (multi--shift CG is used for all the factors in the rational approximation). See table \ref{tbl:qcdqedruns} for details on the degrees and spectral ranges of the rational approximation used for the $(D^\dag D)^{-1/4}$ (up quark) and $(D^\dag D)^{-1/2}$ (down and strange quarks).

The SU(3) and U(1) action densities are defined as:
\begin{eqnarray}
E_\text{SU(3)}(t)&=&\frac{1}{2L^3T} \sum_{{x}} \sum_{\mu\nu} \tr \{ \widehat{F}_{\mu\nu}(x,t) \widehat{F}_{\mu\nu}(x,t) \} \ , \label{eq:u1flow} \\
E_\text{U(1)}(t)&=&\frac{1}{2L^3T} \sum_{{x}} \sum_{\mu\nu} \tr \{ \widehat{A}_{\mu\nu}(x,t) \widehat{A}_{\mu\nu}(x,t) \label{eq:su3flow} \}
\ ,
\end{eqnarray}
where $\widehat{F}_{\mu\nu}(x,t)$ and $\widehat{A}_{\mu\nu}(x,t)$ are the clover--type discretization of the SU(3) and U(1) field strength tensors respectively, at positive flow time. The global topological charge for SU(3) gauge fields is given by:
\begin{eqnarray}
Q_\text{top}(t)&=&\frac{1}{32\pi^2} \sum_{x} \sum_{\mu\nu\rho\sigma} \epsilon_{\mu\nu\rho\sigma} \tr \{ \widehat{F}_{\mu\nu}(x,t) \widehat{F}_{\rho\sigma}(x,t) \}
\ . 
\label{eq:qtop}
\end{eqnarray}

\bibliography{lattice2017}

\end{document}